\documentstyle[amssymb,preprint,aps]{revtex}

\begin{document}
\title{Disorder and Interaction in 2D: Exact diagonalization study of the
Anderson-Hubbard-Mott model}
\author{R. Kotlyar$^{(1)}$ and S. Das Sarma$^{(2)}$}
\address{$^{(1)\text{ }}$Naval Research Laboratory, Washington, D.C. 20375\\
$^{(2)\text{ }}$Department of Physics, University of Maryland, College Park,%
\\
MD 20742}
\date{\today}
\maketitle

\begin{abstract}
We investigate, by numerically calculating the charge stiffness, the effects
of random diagonal disorder and electron-electron interaction on the nature
of the ground state in the 2D Hubbard model through the finite size exact
diagonalization technique. By comparing with the corresponding 1D Hubbard
model results and by using heuristic arguments we conclude that it is {\it %
unlikely} that there is a 2D metal-insulator quantum phase transition
although the effect of interaction in some range of parameters is to
substantially enhance the non-interacting charge stiffness.
\end{abstract}


\noindent \qquad Understanding the nature of the ground state in an
interacting disordered electron system is one of the most formidable and
interesting challenges in condensed matter physics. A question of
fundamental importance is whether the ground state of an interacting
disordered electron system is a metal, an insulator, or some other state
(e.g. superconductor). This question takes on particular significance in
two-dimensional (2D) systems where it is generally accepted that (1) the
disordered 2D system in the absence of any interaction is a localized
(weakly localized for weak disorder) insulator, and (2) the interacting
clean 2D system (without any disorder) is a Fermi liquid metal at high
electron densities (and a Wigner crystal at low electron densities). Little
is known about the disordered interacting system when both disorder and
interaction are strong and of comparable magnitudes so that neither may be
treated as a perturbation. A notable attempt \cite{Finkelstein} by
Finkelstein to analytically explore the nature of the disordered interacting
electron system remains inconclusive as the theory flows toward strong
coupling. Recently renewed interest has developed \cite{Abrahams} in this
subject with much of the current motivation arising from a set of
experimental measurements on the low temperature transport properties of low
density 2D electron (or hole) systems confined in Si MOSFETs and GaAs
heterostructures. These transport measurements (carried out as a function of
carrier density) have been interpreted \cite{Abrahams} by many (but {\it not}
all) as exhibiting evidence for a 2D metal-insulator quantum phase
transition (M-I-T) with the system being a metal at high density $n(>n_{c})$
and an insulator at low density $(n<n_{c})$ with $n_{c}$ as the critical
density separating the two phases. If true this would be a striking example
of an interaction driven quantum phase transition since changing density $%
(n) $ is equivalent to tuning the effective ratio of the interaction energy
to the non-interacting kinetic energy of the system. Much interest has
naturally focused on this possible 2D M-I-T quantum phase transition,
particularly because the corresponding non-interacting disordered 2D
electron system is thought on rather firm grounds \cite{Lee} to be always
localized (Anderson localization) and therefore strictly an insulator at T=0
in the thermodynamic limit. (Much of the debate in interpreting the
experimental data relating to the 2D M-I-T phenomenon arises from the fact
that the experiments are necessarily done at finite temperatures and in
finite systems, whereas the theoretical quantum phase transition is a T=0
infinite system phenomenon.) If such a 2D MIT exists it is of great interest
because the metallic phase must be a non-Fermi liquid since it cannot be
adiabatically connected to the corresponding insulating noninteracting 2D
disordered system.

In this letter we address the nature of the ground state of a disordered
interacting 2D electron system numerically by exactly diagonalizing the few
particle 2D interacting Hamiltonian and doing a disorder averaging. We use
the extensively studied 2D Hubbard model \cite{Dagotto} and its natural
extensions for our exact diagonalization calculations. We study the effects
of both on-site (as in the standard Hubbard model) and longer range
interactions whereas the disorder in our model is a random on-site disorder
of strength $W$ (with $W$ denoting the width of the square distribution from
which the on-site disorder energy is randomly chosen). Without interaction,
our model is the 2D Anderson model, which has a localized insulating ground
state, whereas without disorder our model is the Mott-Hubbard model which
has an extended metallic ground state away from half filling. We restrict to
low ``metallic'' filling factors (typically less than quarter filling)
because our central interest is in understanding the continuum systems, and
also because we want to stay below half filling where the 2D Hubbard model
has an interaction driven Mott transition. Our typical exact diagonalization
study uses the Lanczos technique for $N=6$ electrons (with spin) on a $%
4\times 4$ 2D lattice, corresponding to a filling of $\ \nu =6/32=3/16$.
This involves the diagonalization of matrices of $313600^{2}$size. We
typically average over 10 disorder realizations. Following standard
notations \cite{Dagotto} three parameters $t$ (the hopping amplitude), $U$
(the on-site interaction strength), and $W$ (disorder strength) parametrize
our minimal Anderson-Hubbard model. We carry out our exact diagonalization
in the subspace of the total number of electrons $N$ and the total spin
component $S_{z}=[-(N-M)/2+M/2]$ with $M$ being the number of spin up
electrons. We note that the Hilbert space grows exponentially with the
system size, and the results presented in this work are the essential
current limit on what can be achieved via the exact diagonalization
technique for this problem.

To characterize the nature of the ground state, i.e. its localization
properties, we use the technique \cite{Kohn} suggested by Kohn a long time
ago and calculate the charge stiffness $D_{c}$, sometimes also referred to
as the Drude weight, of the finite system. We calculate charge stiffness for
each individual disorder realization exactly through our finite size
diagonalization, and then obtain the root mean square average by averaging
over a number of disorder realizations. The charge stiffness $D_{c}$, which
is simply related to the persistent current \cite{kotlyar}, is the zero
frequency weight of the long wavelength conductance (i.e. the Drude weight)
in the system. As such, it is finite for a metal or a conductor and is zero
for an insulator or a localized system in the thermodynamic limit \cite{Kohn}%
. Thus $D_{c}$ is an eminently reasonable ``order parameter'' for studying
metal-insulator or localization transitions, although the fact that in
finite systems $D_{c}$ must necessarily be finite introduces complications
in interpreting numerical results. We mention that the strict classification
of a metal (finite $D_{c}$) or an insulator ($D_{c}=0$) based on the charge
stiffness applies {\it in the thermodynamic limit} only in the absence of
disorder because diffusive metallic electrons (in the presence of finite
disorder) have algebraically vanishing $D_{c}$ in the thermodynamic limit
(whereas $D_{c}$ vanishes exponentially in the thermodynamic limit for an
insulator). But this is only of academic interest in finite cluster studies
where no strict distinction between metals and insulators exist in any case
since all finite systems, by definition, have finite conductance (and are
therefore ''metals'' in a trivial sense) by virtue of finite size effects.
Charge stiffness (or persistent current magnitude) has been extensively used
in the literature in finite size numerical localization studies (see, for
example, ref. 7 in the context of 2D M-I-T problem) of disordered
interacting systems, and it is empirically well-known that the calculation
of $D_{c}$ in finite systems is an extremely effective way of numerically
studying the localization problem in the presence of both interaction and
disorder. This is mainly because the charge stiffness is closely related to
the phase sensitivity of the system to boundary conditions, which is an
operationally effective way of distinguishing a metal from an insulator. 

In Fig. 1 we show our calculated disorder-averaged charge stiffness for the $%
4\times 4$ 2D Hubbard cluster (with $6$ electrons) as a function of the
on-site repulsion $U$ for various values of the disorder strength $W$. In
the absence of any disorder ($W=0$), the clean 2D Hubbard model away from
half filling is expected to be a metal with a finite value of $D_{c}$,
whereas the corresponding 2D Anderson model ($U=0$, $W\neq 0$) is expected
to be a weakly localized insulator for small $W$ (crossing over to an
exponentially strongly localized insulator for large $W$). The numerical
results for these limiting cases ($W=0$, $U\neq 0$ and $W\neq 0$, $U=0$) are
also shown in Fig. 1 for the sake of comparison and completeness.

The most important generic feature of the results shown in Fig. 1 is the
peak in the charge stiffness at an intermediate value of $U\equiv U_{c}\sim
W $ where the calculated charge stiffness for the finite 2D cluster has a
maximum for a given disorder strength $W$. The charge stiffness $D_{c}$
appears to decrease from this peak value (for a given $W$) for both $%
U\gtrless U_{c}$. Note that $D_{c}$ increases sharply from $U=0$ to $U=U_{c}$%
, and then decreases slowly for $U>U_{c}$. This peak or the maximum in $%
D_{c} $ is rather manifest in Fig. 1 for $W/t=5$ and $3$ (i.e. for strong
disorder) whereas for weak disorder (e.g. $W/t=0.5$ in Fig. 1) the peak
occurs at somewhat larger values of $U/W\gtrsim 2$ and is not so obvious
from Fig. 1 (we have explicitly verified that the peak exists for $W/t=0.5$
also). The actual value of $U_{c}/t$ clearly depends on the disorder
strength $W$, increasing with $W/t$ from $U_{c}/t\simeq 0.95$ for $W/t=0.5$
through $U_{c}/t\simeq 3.0$ for $W/t=3$ to $U_{c}/t\simeq 3.5$ for $W/t=5$.
The qualitative behavior of our results is explained by the competition
between $U$ and $W$ in the Anderson-Hubbard model. In a disordered system
the random potential $W$ favors a maximal occupation (double occupation for
our spin $1/2$ electrons) of the lowest energy sites. The on-site repulsion $%
U$ on the other hand opposes double occupancy and favors configurations with
minimal number of double-occupied sites. This competition between $W$ and $U$%
, where $W$ tends to localize the charge density and $U$ tends to homogenize
the charge density, is a well-known feature \cite{kotlyar} of the disordered
Hubbard model. The results shown in Fig. 1, in particular, the increase of $%
D_{c}$ as $U$ increases from zero for a fixed disorder strength $W$, \ is a
direct result of the competition. Note that the actual crossover behavior of 
$D_{c}(U,W,t)$ shown in Fig. 1 cannot be parametrized by the single
parameter $U/W$ -- $D_{c}$ depends on both $U/t$ and $W/t$. We have carried
out similar calculations in an `extended' Anderson-Hubbard model with a
long-range interaction (in addition to $U$) with results qualitatively
similar to those shown in Fig. 1.

The direct interpretation of our exact finite size results shown in Fig. 1
is that the conductance of a finite disordered 2D system increases when the
interaction is turned on (at a fixed disorder), reaching a maximum for $%
U=U_{c}\sim W$, and then it decreases slowly with still increasing $U$. The
issue of applying these numerical results based on $4\times 4$ 2D clusters
to address the fundamental question of 2D M-I-T is, however, extremely
tricky. For example, one popular recent line of thinking, based mostly on
numerical work involving spinless electrons in finite 2D systems \cite
{Benenti}, has been to interpret equivalent results on interaction-enhanced
conductance as evidence in favor of a 2D M-I-T, with the peak in $D_{c}$ at $%
U\sim U_{c}$ being interpreted as an intermediate metallic phase. We
disagree with this interpretation for reasons to be discussed below. We
emphasize that any conclusion about the existence of a true quantum phase
transition, based entirely on small system numerical results of the type
shown in Fig. 1, is fundamentally problematic since all finite systems
(whether they are conductors or insulators in the thermodynamic limit) have
finite $D_{c}$ in finite size systems. In principle, a finite size scaling
analysis of the numerical results is capable of determining the existence of
a quantum phase transition (i.e. the 2D M-I-T), but in practice, of course,
one does not have anywhere near the number of data points (for various 2D
system sizes) minimally required to carry out a meaningful finite size
scaling analysis in this problem.

Our conclusion that the charge stiffness results depicted in Fig. 1 do not
indicate the existence of a true 2D M-I-T, but instead show a crossover from
an Anderson insulator at small $U/W$ to a disordered Mott insulator (a
``Wigner glass'' phase) at large $U/W$ with an intermediate crossover regime
(around $U=U_{c}\sim W$) of interaction-enhanced finite size conductance (or
equivalently, an enhanced localization length), which is {\it not} a
thermodynamic ``metallic phase'', is based on two complementary sets of
arguments: (1) Comparison with the corresponding one dimensional (1D)
results; and (2) strong circumstantial evidence based on heuristic
theoretical arguments.

To better understand the nature of the 2D disordered Hubbard model we have
carried out an identical finite system charge stiffness calculation on the
corresponding 1D disordered Hubbard model (1D Hubbard rings). We show the
corresponding 1D Anderson-Hubbard model results in Fig. 2 for $6$ electrons
on a $12$ site ring (corresponding to quarter filling). The 1D results of
Fig. 2 are qualitatively identical to the 2D results of Fig. 1: $D_{c}$ in
the disordered 1D Hubbard model initially increases as a function of $U/W$
for a fixed $W$, showing a maximum at $U=U_{c}\sim W$, and then it decreases
slowly for large $U>U_{c}$, exactly as in 2D system. The ``critical'' $%
U_{c}/t$ for the charge stiffness peak in the 1D system is $U_{c}/t\simeq
0.7 $, $3.3$, $4.5$ for $W/t=0.5$, $3$, $5$ respectively (which are not that
different from the corresponding 2D results at $3/16$ filling).

Noting that the charge stiffness results shown in Figs. 1 and 2 in the 2D
and 1D disordered Hubbard models respectively are essentially
indistinguishable (i.e. just by looking at the results of Figs. 1 and 2 one
does not know which one corresponds to 1D and which to 2D since the results
are qualitatively identical) one is forced to conclude that if the results
of Fig. 1 are interpreted as exhibiting evidence for a 2D M-I-T then one
must, based on the results of Fig. 2, infer that there is also a 1D M-I-T in
the disordered 1D Hubbard model as a function of the interaction strength.
We mention in this context that we have verified that the 1D disordered {\it %
extended} Hubbard model (with additional long range interaction) produces
results qualitatively similar to those in the corresponding 2D system --
thus the equivalence between 1D and 2D charge stiffness results is valid for
finite and long range interactions also.

There are, however, very compelling theoretical grounds \cite{giamarchi} to
believe that 1D disordered systems are localized even in the presence of
interaction. Thus, the results of Fig. 2 cannot be interpreted as evidence
for a 1D M-I-T --- instead the maximum in $D_{c}$ as a function of $U$ only
indicates the interaction-induced enhancement of the localization length
(or, equivalently the persistent current \cite{kotlyar}), which in a finite
system, increases the Drude conductance or the charge stiffness. Based on
the striking qualitative similarity between the 1D (Fig. 2) and the 2D (Fig.
1) results and the fact that both systems have strictly localized or
insulating ground states in the disordered, $W\neq 0$, non-interacting, $U=0$%
, system) we therefore conclude that the 2D results of Fig.1 do not indicate
a 2D M-I-T; it only indicates an interaction-induced enhancement of the 2D
localization length for intermediate interaction strengths $U\sim U_{c}$.
Note that while the intermediate-interaction crossover regime ($U\sim U_{c}$%
) is {\it not} a new quantum phase (it is still an insulator), the
interaction-induced enhancement of the 2D localization length may be
extremely large, and even the experimental 2D systems \cite{Abrahams}
showing the so-called 2D M-I-T may actually be ``effective'' metals since
the enhanced localization lengths may be larger than the actual system size
(or, the phase breaking length at finite temperatures).

In addition to the above empirical argument for the non-existence of a 2D
M-I-T based on the comparison between 1D and 2D exact diagonalization
results we have a heuristic theoretical argument which points to the same
conclusion. The small $U(\rightarrow 0)$ and the large $U(\rightarrow \infty
)$ interaction limits of the disordered 2D Hubbard model are believed to be
insulating or localized on theoretical grounds. The non-interacting $%
(U\rightarrow 0)$ disordered 2D system is known to be localized for any
finite disorder (the localization length is exponentially large, the
so-called weak localization regime, for small disorder) by virtue of the
scaling theory of localization \cite{Lee}. The localized large $%
U(\rightarrow \infty )$ regime arises from the fact that the pure Hubbard
ground state (in the absence of disorder) must have strong ferromagnetic
correlations in the large-$U$ limit in order to minimize the interaction
energy. In fact, it is known \cite{Hlublina} that the large $U$ ground state
of a Hubbard-type model with an additional next-nearest neighbor hopping
term is ferromagnetic (the same is true for the pure Hubbard model at
fillings close to half). In this limit, therefore interaction tends to
become less relevant since the electrons being spin polarized avoid each
other. The system in this large-$U$ limit may thus be approximately
equivalent to a non-interacting or weakly interacting system (albeit a
spin-polarized one), and the introduction of any disorder ($W\neq 0$)
necessarily localizes this 2D ``effectively non-interacting'' Hubbard
system. The weakly localized (for small disorder) large $U(\rightarrow
\infty )$ 2D system has, however, an exponentially longer localization
length (which explains the enhanced $D_{c}$ for large $U$ in Fig. 1) than
the usual non-interacting ($U\rightarrow 0$) disordered limit because the
ferromagnetic spin-polarized phase ($U\rightarrow \infty $) has a larger
Fermi energy, which would exponentially enhance the localization lengths.
Thus, both the small $U$ and the large $U$ regimes are necessarily
localized, and the enhancement of $D_{c}$ in the intermediate-$U(\sim U_{c})$
regime must either indicate a crossover between an Anderson insulator ($%
U\sim 0$) and a disordered Mott insulator (equivalently a Mott glass,
``Wigner glass'' in the corresponding continuum system) for $1/U\sim 0$ or
involve {\it two} quantum phase transitions -- one from the low-$U$ Anderson
insulator phase to the intermediate ($U\sim U_{c}$) ``metallic'' phase with
enhanced $D_{c}$ and then again from this intermediate ``metallic'' phase to
the large-$U$ Mott glass phase. We see absolutely no features in our 2D or
1D numerical results which could be indicative of such a double or
re-entrant insulator ($U\sim 0$) - ``metal'' ($U\sim U_{c}$) - insulator ($%
1/U\sim 0$) quantum phase transition.

We conclude with a critical discussion of the recent low temperature
experimental results in low density, high mobility 2D systems which have
motivated the current resurgence in the issue of 2D M-I-T in disordered and
interacting electron systems. Experimentally one finds \cite{Abrahams} that
the high density regime ($n>n_{c}$) is ``metallic'' in the sense of having a
positive temperature coefficient ($\frac{d\rho }{dT}>0$) of the resistivity $%
\rho $ and the low density ($n<n_{c}$) is insulating with $\frac{d\rho }{dT}%
<0$. This has been interpreted by many \cite{Abrahams} (but not all \cite
{simmons}-\cite{sds}) as clear evidence of an interaction-driven M-I-T
occurring at a critical density $n_{c\text{ }}$.\ The standard
interpretation of these experimental observations as a 2D M-I-T is, however,
problematic because the high density phase (i.e. the {\it less} interacting
phase) is the nominal ``metallic'' phase according to this interpretation.
This makes little sense since the non-interacting or the weakly interacting
very-high density phase must be a weakly localized 2D insulator based on the
scaling theory \cite{Lee}. Thus, very similar to the conclusion we reached
for our exact diagonalization numerical results, the experimental situation
must correspond to either a double quantum phase transition (the very high
density phase is a weakly localized insulator, with the intermediate regime,
corresponding to our peak in $D_{c}$ around $U\sim U_{c}$, being a novel
interaction-induced ``metallic'' phase) or just a sharp crossover from a
high density weakly localized insulator to a low density strongly localized
insulator occurring around $n\sim n_{c}$. Logically, there can be either two
quantum phase transitions (insulator$\rightarrow $metal$\rightarrow $%
insulator) or none, based on our knowledge that the asymptotic high and low
density phases are both insulating phases with the high density phase being
the standard weak localized phase and the low density phase being a strongly
localized phase. Experimentally, there is little evidence for {\it two}
quantum phase transitions (note that there must be two quantum phase
transitions or none; it cannot be one quantum phase transition and one
crossover). Therefore we believe, based on arguments similar to what we use
to interpret our theoretical results presented in this paper, that the
experimental observations are indicating a very sharp crossover (around $%
n\sim n_{c}$) from a weakly to a strongly localized 2D insulator as $n$
decreases, and the high density regime ($n>n_{c}$) is only an effective
``metal'' because the effective system size (the phase breaking length at
finite $T$) is smaller than the localization length which may have been
substantially enhanced by interaction effects as we show in this paper.
There is some very recent experimental support \cite{simmons} for this
scenario.

We emphasize that the interaction induced enhancement of $D_{c}$ for $%
0<U/W\lesssim 1$ in Fig. 1 should not be considered as evidence in favor of
a 2D M-I-T (as was recently done in ref. 7 based on finite system studies of 
{\it spinless} electrons using smaller system sizes) particularly since (1)
the interaction enhancement is only effective for very large disorder
strength ($W/t>1$) where the system is likely to be localized any way (note
that for weak disorder, $W/t<1$, there is essentially no interaction induced 
$D_{c}$ enhancement -- if there is indeed an interaction-driven 2D metallic
phase it is likely to be in the {\it low} disorder regime where Fig. 1
indicates little interaction enhancement), and (2) the actual
interaction-enhanced $D_{c}$ values in the strong disorder regime in Fig. 1
are still extremely small in magnitude (and are much smaller than the
corresponding $D_{c}$ values for non-interacting weak disorder system, which
is still known to be weakly localized by virtue of scaling localization). We
therefore conclude that the interaction enhancement of $D_{c}$ seen in Fig.
1 (and 2) indicates an interaction-driven enhancement of the localization
length in the strong disorder regime, and {\it not} a 2D M-I-T.

We thank Eugene Demler for stimulating discussions. We also acknowledge
helpful correspondence with Andy Millis, Charles Stafford and Dieter
Vollhardt. This work is supported by the US-ONR.

\begin{figure}
\caption{The RMS (averaged over 10 disorder realizations)
charge stiffness $D_{c}$ as a function of onsite repulsion $U$ in 
the 2D $4 \times 4$ disordered Hubbard cluster for $6$ electrons. 
Results for four values of disorder ($W/t=$5, 3, .5, 0) are shown with
the abscissa for the clean ($W=0$) system in the top. Inset: shows 
$D_{c}$ for $W/t=0.5$ in an expanded scale.}
\label{f1}
\end{figure}%
%

\bigskip

\begin{figure}
\caption{The same as in Fig. 1 for the 1D disordered Hubbard ring of $12$ sites 
and $6$ electrons.}
\label{f2}
\end{figure}%
%

\end{document}